\documentclass{jetpl}
\usepackage{epsfig,cite}
\DeclareMathOperator{\Tr}{Tr} \DeclareMathOperator{\Imm}{Im}
\DeclareMathOperator{\sign}{sign}\twocolumn
\lat

\title{ Transitions between``$\pi$'' and ``$0$'' states  in superconductor -- ferromagnet -- superconductor junctions. }

\rtitle{Transitions between``$\pi$'' and ``$0$'' states  in superconductor -- ferromagnet --
superconductor junctions. }

\author{N.\,M.~Chtchelkatchev\/\thanks{e-mail: nms@itp.ac.ru},}
\rauthor{N.\,M.~Chtchelkatchev }

\sodauthor{N.\,M.~Chtchelkatchev}

\address{L.\,D.~Landau Institute for Theoretical Physics RAS,
117940 Moscow, Russia\\
 Institute for High Pressure Physics, Russian Academy of Sciences, Troitsk 142092,
Moscow Region, Russia\\
Moscow Institute of Physics and Technology,  Moscow 141700, Russia}
\dates{\today}*

\abstract{Experimental and theoretical study  of superconductor (S) -- ferromagnet (F) --
superconductor  junctions showed that in certain range of parameters (e.g., the length of the
ferromagnet $d_F$, the exchange field, $E_{\rm ex}$) the ground state of a SFS junction
corresponds to superconducting phase difference $\pi$ or $0$. The phase diagram of a SFS
junction with the respect to $\pi$ and $0$ states is investigated in this letter in $E_{\rm
ex}, d_{F}, T$ space. It is shown that the phase diagram is very sensitive to the geometry of
the system, in particular, to the amount of disorder disorder. }


\begin{document}
\maketitle

Recently many interesting phenomena were investigated in Superconductor (S) - Ferromagnet (F)
- Superconductor Josephson contacts. One of the most interesting effects is the so called
$\pi$-state of SFS junctions
\cite{pi_1,pi_2,Buzdin,Ryazanov,Ryazanov_new,Kontos,Fogelstrom,Nikolai-SFSpi,Barash,Radovic,Golubov_Kupriyanov_Fominov}
in which the equilibrium ground state is characterized by an intrinsic phase difference $\pi$
between the two superconductors.

Theoretical study of SFS junctions  \cite{pi_1,pi_2,Buzdin} showed that if $E_{\rm ex}$ is
fixed in the ferromagnet then the $\pi$-state appears at $0<d_F^{(1)}<d_F<d_F^{(2)}$,
$d_F^{(2)}<d_F^{(3)}<d_F<d_F^{(4)}$ \textit{ etc}..., and near $d_{F}^{(i)}$ ($i=1,2,\ldots$)
the critical current $I_c(d_F)$ has a cusp. Recent experiments \cite{Ryazanov,Ryazanov_new}
showed that the critical current -- temperature curve, $I_c(T)$, in SFS junctions at
$d_F\approx d_{F}^{(i)}$, $i=1,2$ has also a cusp. The temperature of the cusp was identified
with $\pi-0$ transition temperature. These experiments became the motivation of the
theoretical investigations of $I_c(T)$ curves and phase diagrams of SFS junctions. It was
shown in Refs. \cite{Nikolai-SFSpi,Barash} that if $d_F\approx d_{F}^{(i)}$ ($i=1,2,\ldots$)
in very short \textit{ballistic} SFS junctions ($d_F\ll \xi_0=\hbar v_F/\Delta(T=0)$) then
there is $\pi-0$ transition at certain temperature and the $\pi$ state is always (for all $i$)
at larger temperatures then $0$-state. This prediction is in contradiction with the
experimental data and calculations of the phase diagram in dirty SFS junctions based on
linearized Usadel equations, see Ref.\cite{Ryazanov_new} and Refs. there in, where the order
of $\pi$ and $0$ phases with the respect to the temperature is one at $d_F\approx d_{F}^{(1)}$
and the opposite at $d_F\approx d_{F}^{(2)}$. SFS junctions investigated in
Ref.\cite{Ryazanov_new} were dirty. From the first glance it may seem that disorder strongly
influences on the phase diagram of SFS junctions. This is exactly so. It is shown in this
letter that the phase diagram of SFS junctions is rather sensitive the geometry of the system,
in particular, to the amount of disorder in the junction.

The paper is organized as follows: most of the paper is devoted to investigations of phase
diagrams of SFS junctions near the first cusp of $I_c(d_F)$ (at $d_F\approx d_{F}^{(1)}$), and
the case $d_F\approx d_{F}^{(i)}$, $i>1$ is discussed in end.

\begin{figure}[t]
\begin{center}
\includegraphics[height=50mm]{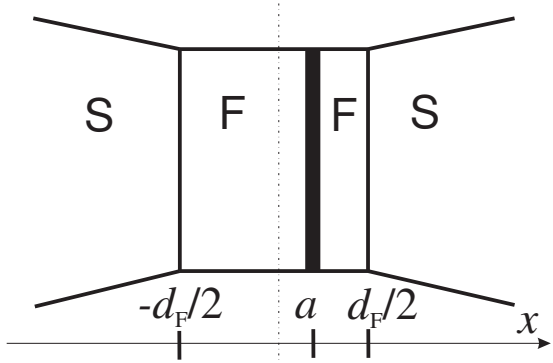}
\caption{\label{fig1} Fig.1. A sketch of a SFS junction. A  barrier (e.g., an insulator layer)
is situated at position $x=a$ from the junction center.}
\end{center}
\end{figure}

The superconducting SFS junction that is investigated here is sketched in Fig.\ref{fig1}. A
barrier (e.g., an insulator layer) is situated at position $x=a$ from the junction center.

First will be considered ballistic SFS junctions. I assume that the exchange energy of the
ferromagnet $E_{\rm ex}\ll E_F$; there is no barrier at SF boundaries: the probability of
Andreev reflection of subgapped Bogoliubov quasiparticles from a SF boundary is equal to
unity; variation of the superconducting gap in S near the boundary will be neglected (this is
correct approximation if our SFS is a quantum point contact \cite{Beenakker,Furusaki}; in
other cases this approximation can be used because it usually leads only to few percent
mistakes in the Josephson current).
Then the Josephson current can be found, for instance, using scattering matrix method
\cite{Beenakker,Nikolai-SFSpi}:
\begin{gather}\label{eq_I}
I(\varphi)=\frac{2e}{\hbar}\frac 1 2\sum_{\sigma=\pm 1}T\sum_\omega \partial_\varphi\ln
g(\omega,\varphi,\sigma),
\end{gather}
where
\begin{multline}\label{eq_g}
g(\omega,\varphi,\sigma)=(2\omega^2+\Delta^2)\cosh(\Phi)+
\\
+2\omega\sqrt{\omega^2+\Delta^2}\sinh(\Phi)+\cal T\cos\varphi+\cal R\cosh\beta.
\end{multline}
Here $\varphi$ is the phase difference between the superconductors; $\Delta$ is the bulk
superconducting gap; $\omega=2\pi(n+1/2)$, $n=0,\pm 1,\ldots$; $\Phi=2d_F(\omega+iE_{\rm
ex}\sigma)/\hbar v_F\cos\theta$, $\beta=4a(\omega+iE_{\rm ex}\sigma)/\hbar v_F\cos\theta$ and
${\cal T}=t_\uparrow t_\downarrow$, ${\cal R}=r_\uparrow r_\downarrow$, where
$(t_\uparrow)^2$, $(r_\uparrow)^2$ are transmission and reflection   probabilities of the
barrier for spin-up electrons, and $\theta$ is the angle between the trajectory and the
$X$-axis. Eqs.\eqref{eq_I}-\eqref{eq_g} can be generalized if the ratio $E_{\rm ex}/E_{F}$ is
arbitrary. Then, for example, $\Phi=\Imm d_F\{\sqrt{2m(E_F+i\omega+E_{\rm
ex}\sigma)}-\sqrt{2m(E_F-i\omega-E_{\rm ex}\sigma)}\}$, $\beta$ can be written in a similar
way. $\omega$-dependence of $\Phi$ was neglected in
Refs.\cite{Fogelstrom,Nikolai-SFSpi,Barash} because there $d_F$ was much smaller then $\xi_0$.
Physically ballistic model of a SFS junction could be realized, for example, in gated
heterostructures\cite{Takayanagi} or in the break-junctions\cite{Scheer} in external magnetic
field producing Zeeman splitting of Andreev levels.\cite{Yip}

The case $a=0$ (then the scattering potential of the junction is symmetric),
$d_F\lesssim\xi=\hbar v_F/\Delta$ was considered in the papers
\cite{Fogelstrom,Barash,Nikolai-SFSpi}. This model is important because on qualitative level
it well describes SFS junctions where F is a ferromagnetic granule or a spin-active interface,
see Ref.\cite{Fogelstrom} and refs. therein. It was shown that if one fixes $E_{\rm ex} $ and
changes the temperature then the $\pi$-state of a ballistic SFS junction appears usually at
higher temperatures then $0$-state. This is not so if the junction is dirty and there is no
other scattering potential in the junction then the impurity potential. To illustrate this
point of view i consider Josephson current in two similar SFS junctions with the same
dimensionless normal conductance per channel,  same $\Delta$, $E_{\rm ex}$, number of open
channels, etc... The only difference between the junctions that in the first one the normal
conductance is provided by an insulator layer with a flat surface at the center of the
ferromagnet like in Fig.\ref{fig1} and in the second one
--- by nonmagnetic impurities in the ferromagnetic region.
\begin{figure}[t]
\begin{center}
\includegraphics[height=105mm]{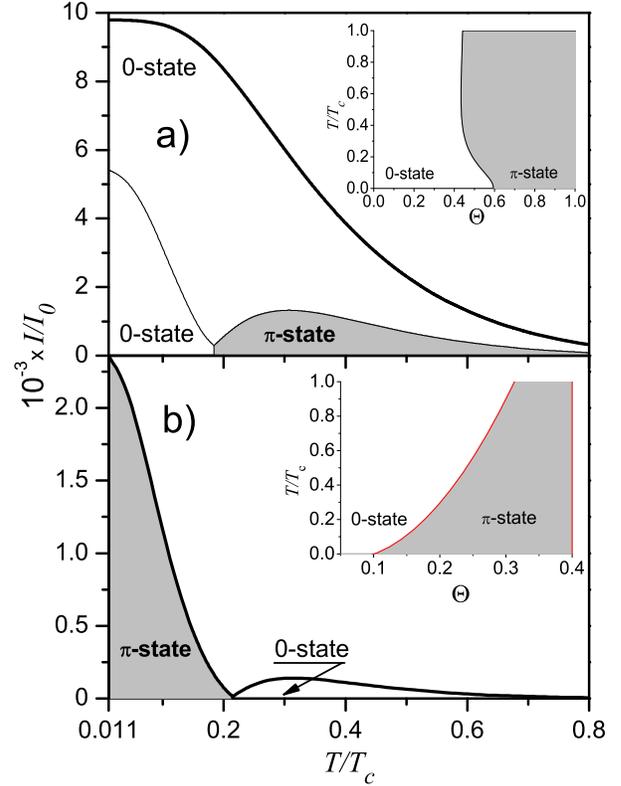}
\caption{Fig.2. The figures show critical current -- temperature relations in two short SFS
junctions with the same normal conductance, near the first minimum of $I_c(d_F)$ (at
$d_F\approx d_F^{(1)}$). Fig.\ref{fig2}a corresponds to ballistic junction and Fig.\ref{fig2}b
--- to dirty one. The insets show ``phase diagrams'' of the junctions. The current scale
$I_0=N_{\rm ch}e\Delta_0/\hbar$, where $N_{\rm ch}$ is the number of open channels in the
junction. The thick curve in Fig.\ref{fig2}a corresponds to the same $\Theta=0.18$ as in
Fig.\ref{fig1}b; the other curve in Fig.\ref{fig2}a corresponds to $\Theta\approx 0.5$. It is
seen that the $\pi$-state in a ballistic SFS junction is at higher temperatures then
$0$-state; the opposite phenomenon takes place in a dirty SFS junction. \label{fig2}}
\end{center}
\end{figure}
Following parameters were chosen: $d_F=\xi_0=\hbar v_F/\Delta_0$, $\Delta_0=\Delta(T=0)$;
$\Theta=2d_F E_{\rm ex}/\pi\xi_0\Delta_0=0.18$. The disorder strength in the second junction
was $\Delta_0/\tau=10$, where $\tau v_F$ is the mean free path. The transmission probability
of the insulator layer in the first junction was
$D(\theta)=D_0\cos^2\theta/(1-D_0+D_0\cos^2\theta)$, with $D_0=0.127$. The current scale
$I_0=N_{\rm ch}e\Delta_0/\hbar$, where $N_{\rm ch}$ is the number of open channels in the
junction. The thick curve in Fig.\ref{fig2}a corresponds to the same $\Theta=0.18$ as in
Fig.\ref{fig1}b; the other curve in Fig.\ref{fig2}a corresponds to $\Theta\approx 0.5$. It
follows that near the first cusp of $I_c(d_F)$ (i.e., $d_F\sim d_F^{(1)}$) $\pi$-state in a
short ballistic SFS junction is at higher temperatury then $0$-state; the opposite phenomenon
takes place in a dirty SFS junction. If there is an insulator barrier in a ballistic short SFS
junction and we add nonmagnetic impurities in it then the transition from the case shown in
Fig.\ref{fig2}a to the case shown in Fig.\ref{fig2}b is expected to occur when $l/d_F\sim D$,
where $l$ is a mean free path. Josephson current calculations in ballistic junctions were
performed using Eqs.\eqref{eq_I}-\eqref{eq_g} with $t_\uparrow=t_\downarrow=\sqrt D$, in dirty
junctions --- using Ricatti form of Eilenberger quasiclassical equations \cite{Belzig}.
\begin{figure}[t]
\begin{center}
\includegraphics[height=60mm]{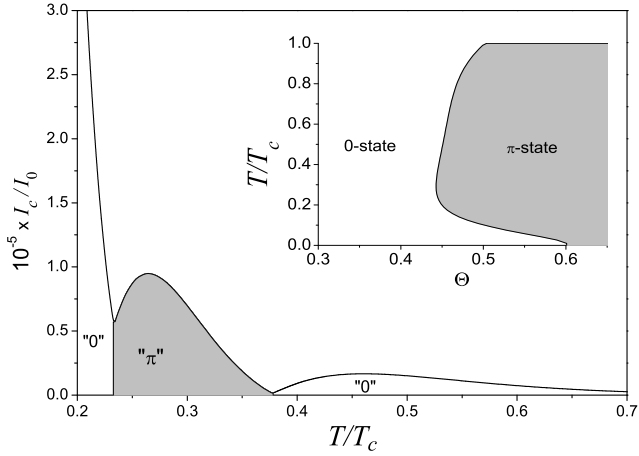}
\caption{Fig.3.The figure exhibits critical current -- temperature relation in a  long
$d_f\approx 3 \xi_0$ ballistic SFS junction with $D_0=0.127$, $\Theta\approx 0.445$, near the
first minimum of $I_c(d_F)$ (at $d_F\approx d_F^{(1)}$). The inset shows the ``phase diagram''
of the junction. If impurities were added in this  type of junctions then the phase diagram
would finally look like in Fig.\ref{fig2}b. If i make $D_0\approx 1$, see Fig.\ref{fig40},
then the phase diagram will become similar to the phase diagram shown in Fig.\ref{fig2}b. }
\label{fig3}
\end{center}
\end{figure}

\begin{figure}[t]
\begin{center}
\includegraphics[height=60mm]{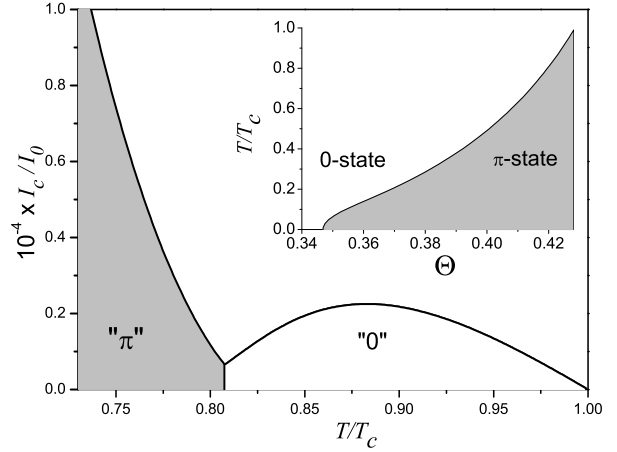}
\caption{Fig.4. The figure shows critical current -- temperature relation in a  long
$d_f\approx 3 \xi_0$ \textit{{ballistic}} SFS junction with $D_0=1$, $\Theta\approx 0.42$,
near the first minimum of $I_c(d_F)$ (at $d_F\approx d_F^{(1)}$). The inset shows the ``phase
diagram'' of the junction. The phase diagram (and the current temperature relation) in a short
dirty SFS junction, Fig.\ref{fig2}b, looks quite similar to one shown in Fig.\ref{fig4}.}
\label{fig40}
\end{center}
\end{figure}

A ``short'' SFS junction with $d_F\lesssim\xi_0$ was considered above. What may happen if
$d_F>\xi_0$ is illustrated in Fig.\ref{fig3}. The junction is ballistic (with the same normal
conductance corresponding to $D_0=0.127$) and $d_F\approx 3\xi_0$. The inset shows the ``phase
diagram'' of the junction. So by freezing long SFS junction (at $d_F\approx d_F^{(1)}$)
starting from $T_c$ one can go through a sequence of phase transitions:
$0\rightarrow\pi\rightarrow 0$. In general one can make junction phase diagram similar to any
phase diagram depicted in Figs.\ref{fig1}-\ref{fig3} by the proper choice of the position of
the insulator barrier and the length of the junction. For example, if i make $D_0= 1$ then the
phase diagram in Fig.\ref{fig1} will transform to the phase diagram shown in Fig.\ref{fig2}b
(the junction is ballistic!). This is shown in Fig.\ref{fig40}. The same transformation of the
phase diagram occurs due to impurities as show numerical calculations.

Below is given a short description of numerical calculation procedure that was used for
drawing of Fig.\ref{fig2}b. Calculations were done using Ricatti representation of the
Eilenberger quasiclassical equations (because Eilenberger equations are unstable)
\cite{Belzig}. The quasiclassical Green functions can be parameterized via the new functions
$a$ and $b$, so
\begin{equation} \label{parametrization}
f=\frac{2a}{1+ab}\sign\omega,\qquad g=\frac{1-ab}{1+ab}\sign\omega.
\end{equation}
The amplitudes $a,b$ change according to the Ricatti equations
\begin{gather}\label{Ricatti}
\mathbf v_F \nabla a + 2\omega_R a +\tilde\Delta_R a^2 -\Delta_R =0,  \\
\mathbf v_F \nabla b - 2\omega_R b -\Delta_R b^2 +\tilde\Delta_R =0, \notag
\end{gather}
where $\Delta_R=\Delta+\frac{\langle f\rangle}{2\tau}$,
$\tilde\Delta_R=\Delta^*+\frac{\langle\tilde f\rangle}{2\tau}$ and $\omega_R=\omega_n+i\sigma
E_{\rm ex}+\langle g\rangle/2\tau$. Here $\langle \ldots \rangle$ denotes averaging over
directions of the quasiclassical trajectories, $\tau$ is the mean free path of an electron in
the impurity potential. Eqs.\eqref{Ricatti} were solved numerically self consistently, the
Josephson current density was evaluated as follows:
\begin{gather}
j=-i\pi e \nu T\sum_\omega\sum_\sigma\langle \mathbf v_F g \rangle_{\mathbf v_F},
\end{gather}
where $\nu$ is the normal density of states. I tried to expand the Ricatti equations over
$\tau^{-1}$ in the first order to find analytically how disorder influences on the phase
diagram of a short SFS junction. Calculations showed no effect in the first order.

The discussion above may lead to a conclusion that in \textit{dirty} Josephson junctions with
ferromagnet layers between superconductors the phase diagram always looks similar like in
Fig.\ref{fig2}b if $d_F\sim d_F^{(1)}$. It seems that this statement is true for short SFS
junctions, but this is not true in general. To prove it i shall give below an example (more
examples and detailed discussion will be give in the extended version of this letter).
Consider a dirty SFIFS junction (e.g., like in Fig.\ref{fig1}a). The Josephson current can be
found from Usadel equations \cite{Belzig}. In general Usadel equations are nonlinear over
quasiclassical greens functions. But near the critical temperature of the junction or if
superconductors and ferromagnets are weakly coupled the Usadel equations can be linearized.
Then   the linearized Usadel equations in the ferromagnets for anomalous function $f$ look
like:
\begin{gather}
\partial^2 f-\frac {2|\omega|+2i{J_{1,2}}\sigma \sign\omega} {D_{1,2}} f=0,
\end{gather}
where $D_{1,2}$ is the diffusion constant and $J_{1,2}$ --- the exchange energy in the left
(right) ferromagnetic layer. If the magnetizations of the ferromagnetic layers are collinear
then the $f$-function to the left of the insulator layer I, see Fig.\ref{fig1}b, is $f=\frac 1
{\sqrt{|\kappa_1/\rho_1|}}(Ae^{\kappa_1 x}+Be^{-\kappa_1 x})$ and to the left
--- $f=\frac 1 {\sqrt{|\kappa_2/\rho_2|}}(Ce^{\kappa_2 x}+De^{-\kappa_2 x})$. Here
$\kappa_{1,2}=\sqrt{(|\omega|+2i{J_{1,2}}\sigma \sign\omega)/ D_{1,2}}$, where $\rho_{1,2}$ is
the resistivity of the ferromagnets. The amplitudes $A,B,\ldots$ are not independent but they
are connected by boundary conditions \cite{Kuprianov-lukichev} that I write here in the matrix
form:
\begin{gather}\label{Eq_BCAD}
    \begin{pmatrix}
      B \\
      C \\
    \end{pmatrix}=S
    \begin{pmatrix}
      A \\
      D \\
    \end{pmatrix},
    \\
    S=\begin{pmatrix}
      e^{2\kappa_1 a}\frac{\tilde{\kappa}_1-\tilde{\kappa}_2-\tilde{\kappa}_1\tilde{\kappa}_2R}
      {\tilde{\kappa}_1+\tilde{\kappa}_2-\tilde{\kappa}_1\tilde{\kappa}_2R} &
      e^{(\kappa_1-\kappa_2) a}\frac{\sqrt{\tilde{\kappa}_1\tilde{\kappa}_2}}
      {\tilde{\kappa}_1+\tilde{\kappa}_2-\tilde{\kappa}_1\tilde{\kappa}_2R}
      \\\\
      e^{(\kappa_1-\kappa_2) a}\frac{\sqrt{\tilde{\kappa}_1\tilde{\kappa}_2}}
      {\tilde{\kappa}_2+\tilde{\kappa}_1-\tilde{\kappa}_1\tilde{\kappa}_2R} & e^{-2\kappa_2 a}\frac{\tilde{\kappa}_2-\tilde{\kappa}_1-\tilde{\kappa}_1\tilde{\kappa}_2R}
      {\tilde{\kappa}_2+\tilde{\kappa}_1-\tilde{\kappa}_1\tilde{\kappa}_2R} \\
    \end{pmatrix},
\end{gather}
where $S$ is the ``scattering matrix'' of the F-F boundary (diagonal elements of $S$ play the
role of ``reflection'' amplitudes, off diagonal --- ``transmission'' amplitudes).
\begin{figure}[t]
\begin{center}
\includegraphics[height=60mm]{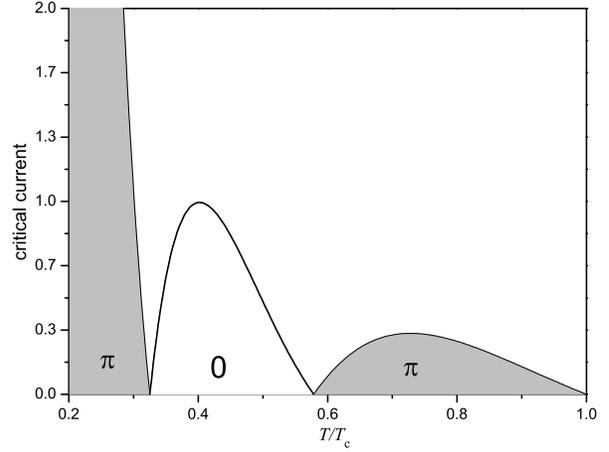}
\caption{Fig.5. The figure shows critical current -- temperature dependence in the dirty
SFIF'S junction with parameters: $J_1=-1.9T_c$, $J_2=3T_c$, $d_F/\xi=1$, $a/\xi=-0.2$,
$D_1=D_2$, ($\xi=\sqrt{D/T_c}$), $R/(\rho_1\xi)=0.1$. Here $d_F$ is again near $d_F^{(1)}$.
The critical current is normalized on its value in the maximum of $I_c(T)$ corresponding to
$0$-state. } \label{fig4}
\end{center}
\end{figure}
Here $\tilde \kappa_{1,2}=\kappa_{1,2}/\rho_{1,2}$; $R$ is resistance of the F-F boundary.
When $R=0$, the scattering matrix $S$ is similar to the quantum mechanical scattering matrix
of a potential step. At the SF interface the boundary conditions look like:
\begin{gather}
\begin{pmatrix}
  A \\
  D \\
\end{pmatrix}=
S_b\begin{pmatrix}
  B \\
  C \\
\end{pmatrix} -\vec \Delta_{\rm eff},
\\
S_b=\begin{pmatrix}
  e^{2\kappa_1 d_F} & 0 \\
  0 & e^{2\kappa_2 d_F} \\
\end{pmatrix},
\\
\vec \Delta_{\rm eff}=\begin{pmatrix}
  \frac{\Delta_L e^{\kappa_1 d_F}}{\Omega R_L} \\
  \frac{\Delta_R e^{\kappa_2 d_F}}{\Omega R_r} \\
\end{pmatrix},\label{Eq_Delta_eff}
\end{gather}
where $\Delta_{L(R)}=\Delta\exp(\pm i\phi/2)$ are the gaps of the left (right)
superconductors, $\Omega=\sqrt{|\Delta|^2+\omega^2}$, $R_{L(r)}$ are the resistances of the SF
boundaries. From Eqs.\eqref{Eq_BCAD}-\eqref{Eq_Delta_eff} i get:
\begin{gather}
\begin{pmatrix}
  A \\
  D \\
\end{pmatrix}=(S_b S-1)^{-1}\vec \Delta_{\rm eff}.
\end{gather}
The approach applied above is similar with the scattering matrix method used in
Ref.\cite{Beenakker} for calculation of the Josephson current in SNS junctions. Then the
Josephson current density can be found: $j=\frac {\sigma_N\pi
i}{2e}T\sum_\omega\Tr\{\tilde{f}\partial {f}-f\partial \tilde{f}\}$, where $\Tr$ is taken over
spin degrees of freedom and $\tilde f(\omega)=f^*(-\omega)$. Fig.\ref{fig4} shows critical
current -- temperature dependence in the dirty SFIF'S junction with parameters: $J_1=-1.9T_c$,
$J_2=3T_c$, $d_F/\xi=1$, $a/\xi=-0.2$, $D_1=D_2$, ($\xi=\sqrt{D/T_c}$), $R/(\rho_1\xi)=0.1$.
Linearization of the Usadel equations is correct at these parameters. Fig.\ref{fig4} is
similar to Fig.\ref{fig3} corresponding to ballistic SFS junction.
\begin{figure}[t]
\begin{center}
\includegraphics[height=80mm]{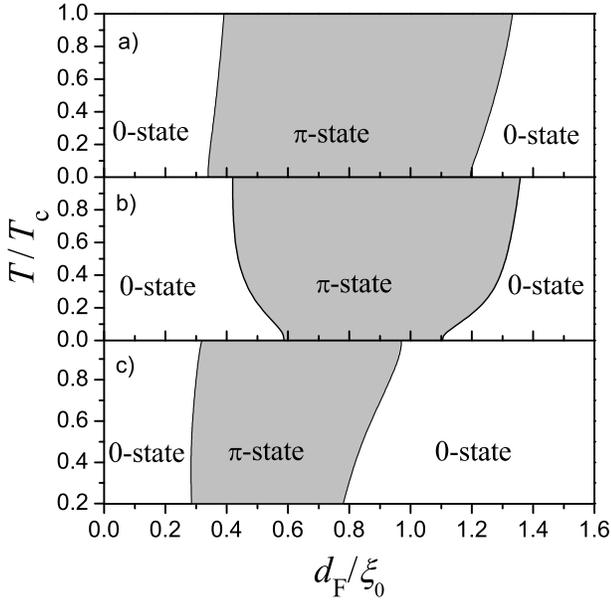}
\caption{Fig.6.  This figure illustrates how geometry of a SFS junction and disorder could
influence on the shape of the phase diagram. In all figures $2E_{\rm ex}/\pi\Delta_0=1$. The
first and the second diagrams, Figs.\ref{fig6}(a-b), correspond to ballistic SFS junctions
described by Eqs.\eqref{eq_I}-\eqref{eq_g} with $D_0=1$ (no layer I, see Fig.\ref{fig1}) and
$D_0=0.127$, $a=0$ correspondingly. The role of impurities can be seen in Fig.\ref{fig6}(c).
There $D_0=1$ and $\hbar/(\Delta_0\tau)=10$. At $d_F/\xi_0=1$ the normal conductance of the
junction in Fig.\ref{fig6}(c) becomes equal to the normal conductance of the junction shown in
Fig.\ref{fig6}(a). If I draw the phase diagrams for the SFS junctions in $\Theta,T_c$ space it
will look similar to Fig.\ref{fig6}.} \label{fig6}
\end{center}
\end{figure}

The phase diagrams were considered above in $(T, \Theta)\sim (T,E_{\rm ex})$ space at fixed
$d_F$. If one fixes $E_{\rm ex}$ near the first cusp of $I_c(E_{\rm ex})$ and changes $d_F$ he
will obtain similar figures.

Everywhere above I considered phase diagrams of SFS junctions near the first cusp of
$I_c(d_F)$ (at $d_F\approx d_{F}^{(1)}$). Below I briefly discuss the phase diagram in general
case. Fig.\ref{fig6} shows tree phase diagrams. In all the figures $2E_{\rm
ex}/\pi\Delta_0=1$. The first and the second diagrams, Figs.\ref{fig6}(a-b), correspond to
ballistic SFS junctions described by Eqs.\eqref{eq_I}-\eqref{eq_g} with $D_0=1$ (no layer I,
see Fig.\ref{fig1}) and $D_0=0.127$ correspondingly. The role of impurities can be seen in
Fig.\ref{fig6}(c). There $D_0=1$ and $\hbar/(\Delta_0\tau)=10$. At $d_F/\xi_0=1$ the normal
conductance of the junction in Fig.\ref{fig6}(c) becomes equal to the normal conductance of
the junction shown in Fig.\ref{fig6}(a). If I'd use the linearized Usadel equations to
describe the phase diagram of a SFS junction then I'd get the graph like in Fig.\ref{fig6}(c).
The phase diagram in Fig.\ref{fig6}(c) qualitatively agrees with experimental results in
Ref.\cite{Ryazanov_new} that the order of $\pi$ and $0$ phases with the respect to the
temperature is one at $d_F\approx d_{F}^{(1)}$ and the opposite at $d_F\approx d_{F}^{(2)}$.

In conclusion, $\pi-0$ transitions in Josephson junctions with ferromagnetic layers are
investigated in this letter. It is shown that the phase diagram is very sensitive to the
geometry of the system, in particular, to the amount of disorder in the junction.

I'm grateful to I.S. Burmistrov, A.S. Iosselevich and especially Yu.S. Barash for stimulating
discussions. I also thank RFBR ~Project No. 03-02-16677 and No 02-02-16622, the Russian
Ministry of Science, the Netherlands Organization for Scientific Research ~NWO, CRDF, Russian
Science Support foundation and State Scientist Support foundation (Project No. 4611.2004.2).


\begin{thebibliography}{99}
\bibitem{pi_1}L.\,N.\,Bulaevskii, V.\,V.\,Kuzii, and A.\,A.\,Sobyanin,
Pis'ma Zh. Eksp. Teor. Fiz. {\bf 25}, 314 (1977) [JETP Lett. {\bf 25}, 290 (1977)].

\bibitem{pi_2}A.\,V.\,Andreev, A.\,I.\,Buzdin, and R.\,M.\,Osgood,
Phys. Rev. B {\bf 43}, 10124 (1991).

\bibitem{Buzdin} A.\,I.\,Buzdin, B.\,Vujicic, and M.\,Yu.\,Kupriyanov,
Zh. Eksp. Teor. Fiz. {\bf 101}, 231 (1992) [Sov. Phys. JETP {\bf 74}, 124 (1992)].

\bibitem{Ryazanov} A.\,V.\,Veretennikov, V.\,V.\,Ryazanov, V.\,A.\,Oboznov {\it et al.}, Physica B {\bf 284-288}, 495 (2000); V.\,V. Ryazanov, V.\,A. Oboznov,
A.\,Yu. Rusanov {\it et al.}, Phys. Rev. Lett. {\bf 86}, 2427 (2001).

\bibitem{Ryazanov_new} V.\,V.\,Ryazanov, V.\,A.\,Oboznov, A.S. Prokofiev \textit{ et al}, J.
Of Low Temp. Phys. \textbf{136}, 385 (2004).

\bibitem{Kontos} T. Kontos, M. Aprili, J. Lesueur, \textit{et al}, Phys. Rev. Lett.
\textbf{89}, 137007 (2002)

\bibitem{Fogelstrom} M.\,Fogelstr\"om, Phys. Rev. B {\bf 62}, 11812 (2000).

\bibitem{Nikolai-SFSpi} N.\,M.\,Chtchelkatchev, W.\,Belzig, Yu.\,V.\,Nazarov,
and C. Bruder, JETP Lett., \textbf{74}, 323 (2001) [Pis'ma v Zh. Eksp. i Teor. Fiz., Vol. 74,
No. 6, 2001, pp. 357–361].

\bibitem{Barash} Yu. S. Barash and I. V. Bobkova, Phys. Rev. B \textbf{65}, 144502 (2002).

\bibitem{Radovic} Z. Radovic, N. Lazarides, and N. Flytzanis, Phys. Rev. B \textbf{68}, 014501
(2003).

\bibitem{Golubov_Kupriyanov_Fominov} A.A. Golubov, M.Yu. Kupriyanov, and Ya.V. Fominov,
Pis'maZh. Eksp. Teor. Fiz. \textbf{75}, 223 (2002) [JETP Lett. \textbf{75}, 190 (2002)].

\bibitem{Beenakker} C.W.J.\ Beenakker, Phys.\ Rev.\ Lett.\ {\bf 67},
3836 (1991).

\bibitem{Furusaki} A.\ Furusaki, H.\ Takayanagi, and M.\ Tsukada, Phys.\ Rev.\ Lett.\ {\bf 67}, 132 (1991)
and Phys.\ Rev.\ B {\bf 45}, 10563 (1992).

\bibitem{Takayanagi} H.\ Takayanagi, T.\ Akazaki, and J.\ Nitta,
  Phys.\ Rev.\ Lett.\ {\bf 75}, 3533 (1995).

\bibitem{Scheer} E.\ Scheer, N. Agrait, J.C. Cuevas, A.L.\ Yeyati,
B.\ Ludoph, A.\ Martin-Rodero, G.R.\ Bollinger, J.M.\ van Ruitenbeek, and C. Urbina, Nature
{\bf 394}, 154 (1998).

\bibitem{Yip} S.K. Yip, Phys. Rev. B \textbf{62}, R6127 (2000).



\bibitem{Belzig} W. Belzig, F.~K. Wilhelm, C. Bruder \textit{et
al.}, Superlattices and Microst. \textbf{25}, 1251 (1999).

\bibitem{Kuprianov-lukichev} M.~Yu. Kupriyanov and
V.~F. Lukichev, Zh. Exp. Teor. Fiz. \textbf{94}, 139 (1988) [Sov. Phys. JETP \textbf{67}, 1163
(1988)].



\end{thebibliography}
\end{document}